\documentclass{aa}

\usepackage{multirow}
\usepackage{txfonts}
\usepackage{graphicx}
\usepackage{natbib}
\bibpunct{(}{)}{;}{a}{}{,}

\newcommand{\teff}{$T_{\rm eff}$}
\newcommand{\nhe} {$N$({\rm He})/$N$({\rm H})}
\newcommand{\msun}{$M_{\rm \odot}$}
\newcommand{\rsun}{$R_{\rm \odot}$}

\newcommand{\hd}{\object{HD\,188112}}

\begin{document}
 
\title{Quantitative spectral analysis of the sdB star HD\,188112:\\ A helium-core white dwarf progenitor
}
 
\author{M. Latour\inst{1}, U. Heber\inst{1}, A. Irrgang\inst{1}, V. Schaffenroth\inst{1,2}, S. Geier\inst{1,3}, W. Hillebrandt\inst{4}, F.~K.~R{\"o}pke\inst{5,6},  S. Taubenberger \inst{3,4},  M. Kromer\inst{7}, 
and M. Fink \inst{8}
 }
 
\institute{
 Dr. Karl Remeis-Observatory \& ECAP, Astronomical Institute,
Friedrich-Alexander University Erlangen-Nuremberg, Sternwartstr. 7,
96049 Bamberg, Germany, \email{marilyn.latour@fau.de}
\and Institute for Astro- and Particle Physics, University of Innsbruck,
Technikerstr. 25/8, A-6020 Innsbruck, Austria
\and ESO, Karl-Schwarzschild-Str. 2, 85748 Garching bei M\"{u}nchen, Germany 
\and Max-Planck-Institut f{\"u}r Astrophysik, Karl-Schwarzschild-Str. 1, 85741 Garching bei M{\"u}nchen, Germany
 \and  Heidelberger Institut f{\"u}r Theoretische Studien, Schloss-Wolfsbrunnenweg 35, D-69118 Heidelberg, Germany
 \and Zentrum f{\"u}r Astronomie der Universit{\"a}t Heidelberg, Institut f{\"u}r Theoretische Astrophysik, Philosophenweg 12, D-69120 Heidelberg, Germany
  \and The Oskar Klein Centre \& Department of Astronomy,
  Stockholm University, AlbaNova, SE-106 91 Stockholm, Sweden 
  \and Institut f{\"u}r Theoretische Physik und Astrophysik, Universit{\"a}t W{\"u}rzburg, Emil-Fischer-Str. 31, D-97074 W{\"u}rzburg, Germany
  }

\date{Received 2015 September 25  / Accepted 2015 November 8}

\abstract
{HD 188112 is a bright ($V$ = 10.2 mag) hot subdwarf B (sdB) star with a mass too low to ignite core helium burning and is therefore considered  a pre-extremely low-mass (ELM) white dwarf (WD).  ELM WDs (M $\protect\la$ 0.3 \msun) are He-core objects produced by the evolution of compact binary systems.}
{We present in this paper a detailed abundance analysis of \hd\ based on high-resolution Hubble Space Telescope (HST) near- and far-ultraviolet spectroscopy. We also constrain the mass of the star's companion. }
{We use hybrid non-LTE model atmospheres to fit the observed spectral lines, and to derive the abundances of more than a dozen elements and the rotational broadening of metallic lines. }
{We confirm the previous binary system parameters by combining radial velocities measured in our UV spectra with the previously published values. The system has a period of 0.60658584 days and a WD companion with M $\geq$ 0.70 \msun. By assuming a tidally locked rotation combined with the projected rotational velocity ($v$ sin $i$ = 7.9 $\pm$ 0.3 km s$^{-1}$), we constrain the companion mass to be between 0.9 and 1.3 \msun. We further discuss the future evolution of the system as a potential progenitor of an underluminous type Ia supernova.
We measure abundances for Mg, Al, Si, P, S, Ca, Ti, Cr, Mn, Fe, Ni, and Zn, and for the trans-iron elements Ga, Sn, and Pb. In addition, we derive upper limits for the C, N, O elements and find \hd\ to be strongly depleted in carbon. We find evidence of non-LTE effects on the line strength of some ionic species such as Si~\textsc{ii} and Ni~\textsc{ii}. The metallic abundances indicate that the star is metal-poor, with an abundance pattern most likely produced by diffusion effects.}
{}


\titlerunning{Spectroscopic analysis of HD 188112}
\authorrunning{M. Latour et al.}
\maketitle

\section{Introduction}  

%\setpagewiselinenumbers
%\modulolinenumbers[5]
%\switchlinenumbers
%\linenumbers

HD 188112 is one of the brightest known ($V$=10.2 mag) hot subdwarf B (sdB) stars, and it is a peculiar one. 
Subdwarf B stars are low-mass ($\sim$0.5 \msun) helium-core burning objects with a hydrogen envelope that is not massive enough to sustain significant hydrogen shell burning. These stars form the extreme horizontal branch (EHB), which is a hot extension (\teff\ $\ga$ 21,000 K) of the horizontal branch. The majority of sdB stars are thus found between the zero-age EHB and the terminal-age EHB, which corresponds to the central helium burning phase \citep{heb09}. The subsequent, more rapid evolution during helium shell burning brings the star above the EHB.

Given its fundamental parameters (\teff\ = 21,500 K, log $g$ = 5.66),  \hd\  was classified as an sdB; however, it lies at an odd position in the log $g$--\teff\ diagram, i.e., significantly below the zero-age EHB \citep{heb03}. This is an indication that the star has too little mass to sustain He-core burning. 
The low-mass nature of \hd\ was  confirmed in \citet{heb03} by comparing the trigonometric distance of the star (from the Hipparcos parallax) with the spectroscopic distance;  a mass of $\sim$0.24 \msun\ is required for both values to match. 
Comparing the fundamental parameters of \hd\ with models for post red giant branch evolution \citep{drie98} led to a similar mass of $\sim$0.23 \msun\ (see Figure 4 in \citealt{heb03}). Such a star is likely to be formed via unstable mass transfer during its red giant phase, which leads to the formation of a common envelope. The subsequent spiralling-in of both components, and the eventual ejection of the envelope will leave behind a short-period binary system. If the envelope is removed before the core helium burning begins, the star  evolves with an inert helium core and becomes a helium core white dwarf (WD) \citep{han02,han03}. Such low-mass evolved stars need to be formed via close binary evolution because the universe is not old enough to produce such a star through single-star evolution \citep{mar95}.
It is thus no surprise that \hd\ was found to be radial velocity (RV) variable with an orbital period P = 0.606585 d, a semi-amplitude K$_1$ = 188.3 km s$^{-1}$, a systemic velocity $\gamma_0$ = 26.6 km s$^{-1}$, and a miminum companion mass of 0.73 \msun\ \citep{heb03}.
%Combining the orbital period, the semi-amplitude and the mass of \hd\, they derived a minimum mass of 0.73 \msun\ for the companion.
 This indicates that the companion is a rather massive WD, as a main-sequence (MS) star of such a mass would be detectable in the spectral energy distribution of \hd, which does not show any deviation from that of  a single star.

At the time of the discovery of \hd, very few low-mass He-core objects were known. The first was discovered as the companion of a neutron star in a millisecond pulsar \citep{kerk96};  another serendipitous discovery followed a few years later, among the first data release of the Sloan Digital Sky Survey (SDSS) \citep{lie04}. No more than a handful of such objects had been spectroscopically confirmed \citep{eis06,kawka06,kilic07} when a dedicated search for what are now called extremely low-mass (ELM) WDs was undertaken in the SDSS data base (\citealt{gia15, browr13}, and references therein). Because these stars are the product of close binary evolution, they are monitored for radial velocity variations, and the majority of them have indeed been found to have an orbital period shorter than a day.
By determining the mass of the star and the orbital parameters of the system, it is possible to derive a minimum mass for the companion, which allows to study the future outcome of these systems. Some of them will merge within a Hubble time, and some might be progenitors of type Ia supernovae (SNe Ia). 
Systems with very short periods ($\la$ 0.2 d) are expected to merge within a Hubble time. To date, 38 of these systems have been found (\citealt{gia15,browr13}, and references therein). Their fate depends on the mass of both components. If the mass of the companion is high enough, then it might be a SN Ia progenitor. Once a He-WD has filled its Roche lobe,  a typical scenario involves a C/O core WD companion that starts accreting He-rich material from the He-WD.
Once a sufficiently large shell of He has been accreted, a detonation can be triggered in the He-shell resulting in underluminous type Ia SNe \citep{bild07}. However, some simulations have shown that the explosion of the He-shell almost inevitably triggers a subsequent detonation in the C/O core. Depending on the mass of the C/O core, this might produce SNe Ia explosions \citep{liv90,fink07,fink10,pak13}. This double-detonation scenario happens at a sub-Chandrasekhar mass and a C/O WD accretor of $\sim$1.0 \msun\ is a good candidate for producing a normal SN Ia \citep{sim10}.   
If the mass of the companion is much lower than 1 \msun\ then extreme-helium stars, R Coronae Borealis (RCrB) stars, single hot subdwarfs, or single WDs can ultimately be formed, depending on the mass ratio of the components.
This is why, in order to constrain the future product of the binary systems hosting these ELM WDs, the masses of both components need to be determined accurately.

 The mass of the ELM WD is usually determined by comparing the atmospheric parameters (effective temperature, \teff, and surface gravity, log $g$) of the star with theoretical evolutionary tracks for different masses \citep{drie98, pan07,kilic10,alt13}. However, the precision of such mass determination is limited and also depends on the evolutionary track used. In addition, the mass of the companion depends on the inclination of the system, which remains unknown unless eclipses are seen \citep{stein10,brownwr11_eclipse,ven11,pars11,kil14,hal15} or distortion effects (such as ellipsoidal variations) in the light curve allow  the system's parameters to be constrained \citep{kilic11_lc,her14b}. However, these are very rare cases, and for the vast majority of systems only a minimum companion mass can be determined or, using the mean inclination angle for a random sample,  a most likely companion mass can be estimated. 

Given all of these new ELM systems found (more than 70), \hd\ might look like a common system among others, but this is not the case. First, when compared to the majority of ELM systems, its combination of hot temperature and low surface gravity makes it stand out on the log $g$--\teff\ plane (see, e.g., Fig. 1 in \citealt{browr13}). This indicates that \hd\ is at a rather early stage of its cooling process, which is why it is also classified as a low-mass sdB. Such ``young'' objects are sometimes referred to as pre-ELM WDs. In terms of fundamental parameters, only four systems so far are roughly similar to \hd: KIC 6614501 \citep{sil12}, SDSS J1625+3632 \citep{kilic11}, SDSS J0815+2309 \citep{browr13}, and GALEX J0805-1058 \citep{kawka15}. 
Second, it is by far the brightest of the known ELM systems;   it  has a parallax measured by Hipparcos, and we can expect  a Gaia measurement soon. The parallax allows  a mass to be derived for the star that is independent of evolutionary models.
Additionally, its brightness allows  high-resolution, high signal-to-noise spectroscopy. 

As an alternative to finding eclipses or modeling light-curves of distorted stars, if one assumes the system to be tidally locked, then its inclination can be measured via the rotational broadening ($v_{\rm rot}$ sin $i$) of metallic lines \citep{geier10}. The surface rotation velocity of the star ($v_{\rm rot}$) is estimated from the orbital period (= rotation period) and then, by measuring the rotational broadening, the inclination can be derived. Given the fact that the optical spectrum of the star is devoid of metallic lines, except for a magnesium feature too weak for such a measurement, we turned to the UV range instead where at least some magnesium lines are expected to be observed.
The UV spectra of \hd\ turned out to be much richer in metallic lines than anticipated, allowing an in-depth chemical composition analysis of the star. In our preliminary analysis we presented abundances for Mg, Si, and Fe \citep{lat15}; in this paper, we extend the measured abundances to a dozen additional elements.
This is of great interest since detailed abundance analyses of ELM WD are still scarce \citep{her14,gia14b,kap13}.
We present the UV observations in Sect. 2 and review some of the binary parameters of the system in Sect. 3.  Section 4 is dedicated to the spectroscopic analysis, it includes a description of the models, the determination of the rotational broadening, and the abundances of all metallic species seen in the spectra. Finally, a discussion follows in Sect.~5.

\begin{figure*}[t]
\sidecaption
\includegraphics[width=12cm]{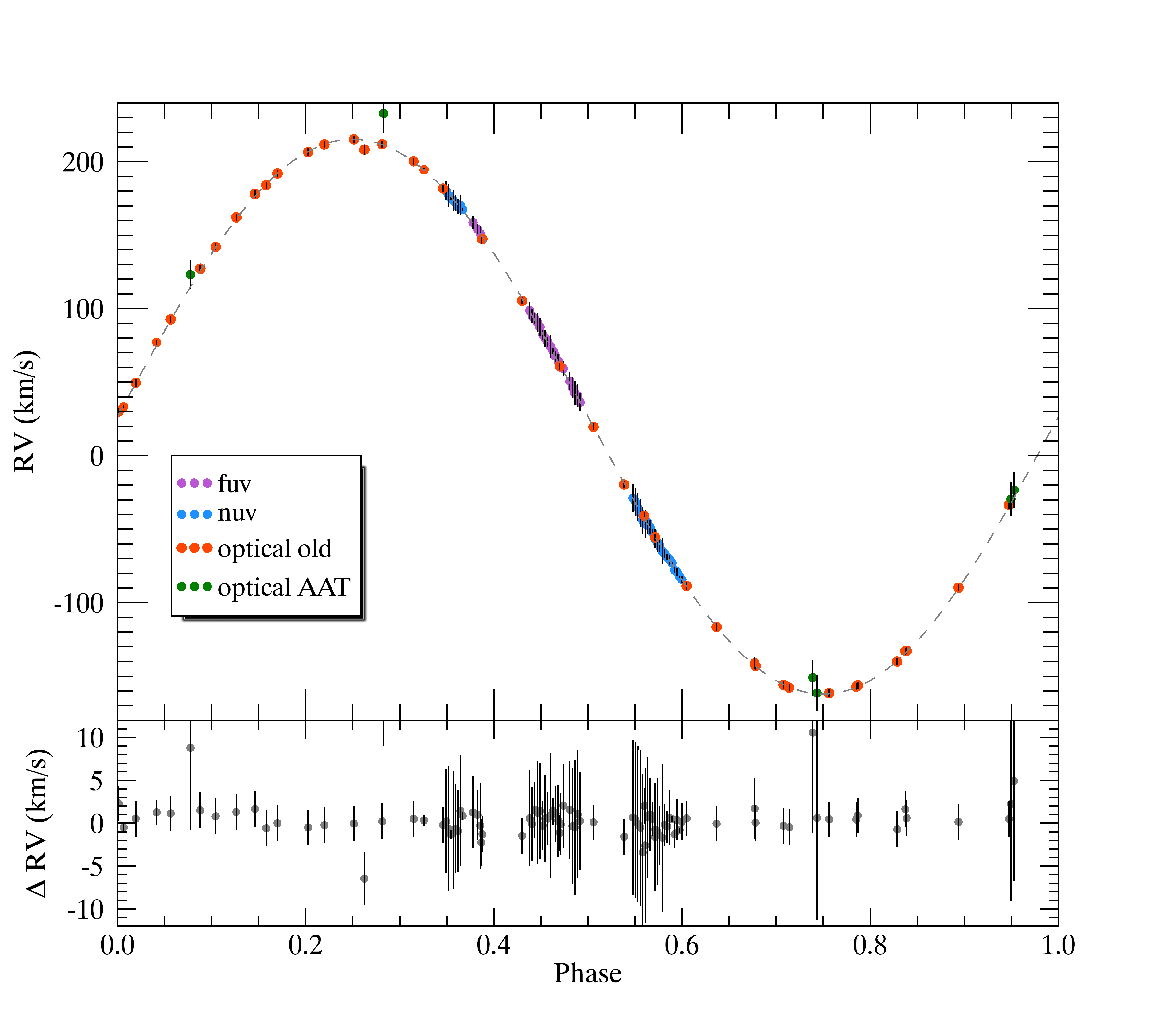}
\caption{Radial velocity curve, with measurements from different sources are indicated with different colors (see Sect. 3). The ``optical old'' refers to values published in \citet{heb03} and \citet{edel05}.  }
\label{rvcurve}
\end{figure*}
\section{Observations} 

Ultraviolet spectra of \hd\ were obtained with the Space Telescope Imaging Spectrograph (STIS) installed on the Hubble Space Telescope (HST) during cycle 20 (program ID 12865) on April 27,$^{\rm }$ 2013. 

First there is a set of 22 exposures of 120 s each taken with the E140H grating (R = 114,000) covering the wavelength range 1242--1440 \AA. 
Because of the radial velocity variations of the star and the high resolution of the observations, a series of 
short exposures was necessary in order to avoid orbital smearing of the metallic lines\footnote{Time-tag mode, as used for the NUV spectrum, would have been preferable, but the star is too bright in the far-UV.}. This exposure time leads to $\sim$3 km s$^{-1}$ differences between subsequent spectra.
The data retrieved from the HST Archive were reduced and processed with the CALSTIS pipeline.
Radial velocities were measured for each exposure and the spectra were shifted to rest wavelength (information on the radial velocity measurements is presented in the following section). 
The 22 ``shifted'' spectra were then coadded into a single spectrum suitable for spectroscopic analysis.

Additionally, two near-UV observations were taken with the E230H grating (also R = 114,000) in Time-Tag mode. The wavelength range covered by these spectra is 2666--2932 \AA. There is a short (1040.2 s, hereafter called NUV1) and a long (2860.2 s, hereafter called NUV2) exposure, taken at different orbital phases. Again, to avoid orbital smearing of the spectral lines, the retrieved data were processed following the STIS Time-Tag Analysis Guide\footnote{http://www.stsci.edu/hst/stis/documents/isrs/200002.pdf} in order to create subexposures that were then reduced using CALSTIS. 
This way, we created 8 spectra of 130 s for the NUV1 observation, and 21 spectra of 136.2 s for the NUV2. 
Radial velocities were measured for each spectrum that was then shifted to rest wavelength. The main spectral features of the near-UV spectra are a series of four Mg~\textsc{ii} lines around 2797 \AA, among which the two resonance lines also have interstellar (IS) components. The NUV1 observation was taken far away from the conjunction phase, so the IS components are distinct from the stellar lines. This is not the case though for the NUV2 observation; the first spectra of the series show blending of the stellar and IS resonance lines.
For this reason, when we coadded the subexposures of NUV2, we included only the 11 spectra where the lines are not blended. As for  NUV1, the 8  subexposures  were combined together.

\section{Binary system parameters}

The HST spectra were obtained for spectroscopic analysis, but the radial velocities measured for the individual exposures are also useful to put new constraints on  the orbital parameters determined in \citet{heb03} and \citet{edel05}, or at least to confirm them. 
By combining our new values with the previous measurements taken between 2000 and 2004, we extended the temporal coverage to 13 years.

Radial velocities were measured in the UV spectra using strong unblended lines: the Si~\textsc{iv} resonance lines (1393.7 and 1402.8 \AA), Si~\textsc{iii} 1296.7 \AA, and Fe~\textsc{iii} 1273.8 \AA.  
For the NUV spectra, the three strongest Mg~\textsc{ii} lines were used, except for six subexposures of NUV2 where the resonance lines were strongly blended with the IS lines. In this case, only the 2798.8 \AA\ line was used.
The mean values of individual measurements were used, which gave maximum standard deviations of 2.5 km s$^{-1}$, but typically of $\sim$1.5 km s$^{-1}$. 
In addition to the previously published optical RVs, we also added five values kindly provided  by J. Norris, from low-resolution spectra taken at the Anglo-Australian Telescope (AAT) in 2003. 

A periodogram analysis was performed, including the 98 RV measurements and our resulting folded RV curve is shown in Fig. \ref{rvcurve}.
The orbital parameters we obtained are in very good agreement with those stated in \citet{heb03}; we have P = 0.60658584 ($\pm$ 5e$-$8) d, K$_1$ = 188.7 $\pm$ 0.2 km s$^{-1}$, and $\gamma_0$ = 26.6 $\pm$ 0.2 km s$^{-1}$. 
This leads to a mass function 
\begin{equation}
f(m)=\frac{M_{\rm{comp}}^{3}\rm{sin^3}\it i}{\left (M_{\rm{comp}}+M\right )^{2}} = \frac{PK^{3}}{2\pi G}
\end{equation}
of $f(m)$ = 0.4221 \msun. 
 By allowing for an eccentric orbit during the fitting procedure, we derive an upper limit of $e \leq$~0.00006.  This low eccentricity value is consistent with a circular orbit.

The parallax of \hd\ from the new reduction of the Hipparcos catalogue is 13.6 $\pm$ 1.7 mas \citep{hip07}. We then used the V (10.22) and B (10.01) magnitudes from the TYCHO 2 catalogue \citep{tycho2} combined with the synthetic flux from a model atmosphere appropriate for \hd\ in order to derive the stellar radius. We did not consider any reddening due to the proximity of the star. Both magnitudes could be satisfactorily reproduced without extinction. This allowed us to derive a spectroscopic mass of 0.245 $^{+0.075}_{-0.055}$ \msun\ and a radius of 0.12 $^{+0.02}_{-0.01}$ \rsun. The main uncertainty results from the parallax measurement. The spectroscopic mass indicates a $M_{\rm comp} \geq$ 0.73 \msun\ and a most likely mass $M_{\rm comp (52\degr)}$ = 1.21 \msun\ \citep{gray08}.

The latest development in terms of evolutionary sequences for ELM WDs is due to \citet{alt13}. Their models predict that for masses in the range 0.18$-$0.4 \msun, the star is expected to undergo H-shell flashes (in CNO burning) during the early cooling phase. This can happen multiple times while the star is ``trying'' to contract and cool down, each time following tracks in a slightly different region of the log $g$--\teff\ diagram (see figs. 3 and 4 of \citealt{alt13}). This makes the determination of the mass a bit more complicated since it is not known if the star is on its final or an intermediate cooling sequence. Moreover, the different cooling tracks of stars from different masses are found in similar places on the HR diagram. The approach of \citet{alt13} is a statistical one, weighting the time spent by models of different masses in each region of the log $g$--\teff\ plane. This way, they determine a mass of 0.211 $\pm$ 0.0175 \msun\ for \hd\, as well as a cooling age of 329 $\pm$ 207 Myr. This mass is a bit lower than that determined by the parallax, but it nevertheless fits well within the spectroscopic mass uncertainty. Assuming 0.21 \msun\ for \hd\ leads to $M_{\rm comp} \geq$ 0.70 \msun\ and a most probable mass $M_{\rm comp (52\degr)}$ = 1.17 \msun. As mentioned before, the absence of NIR excess in the observed flux of the star excludes the possibility of a main-sequence companion, thus implying that it has to be a rather massive WD. 

\begin{figure*}[t!]
\sidecaption
\includegraphics[width=12cm]{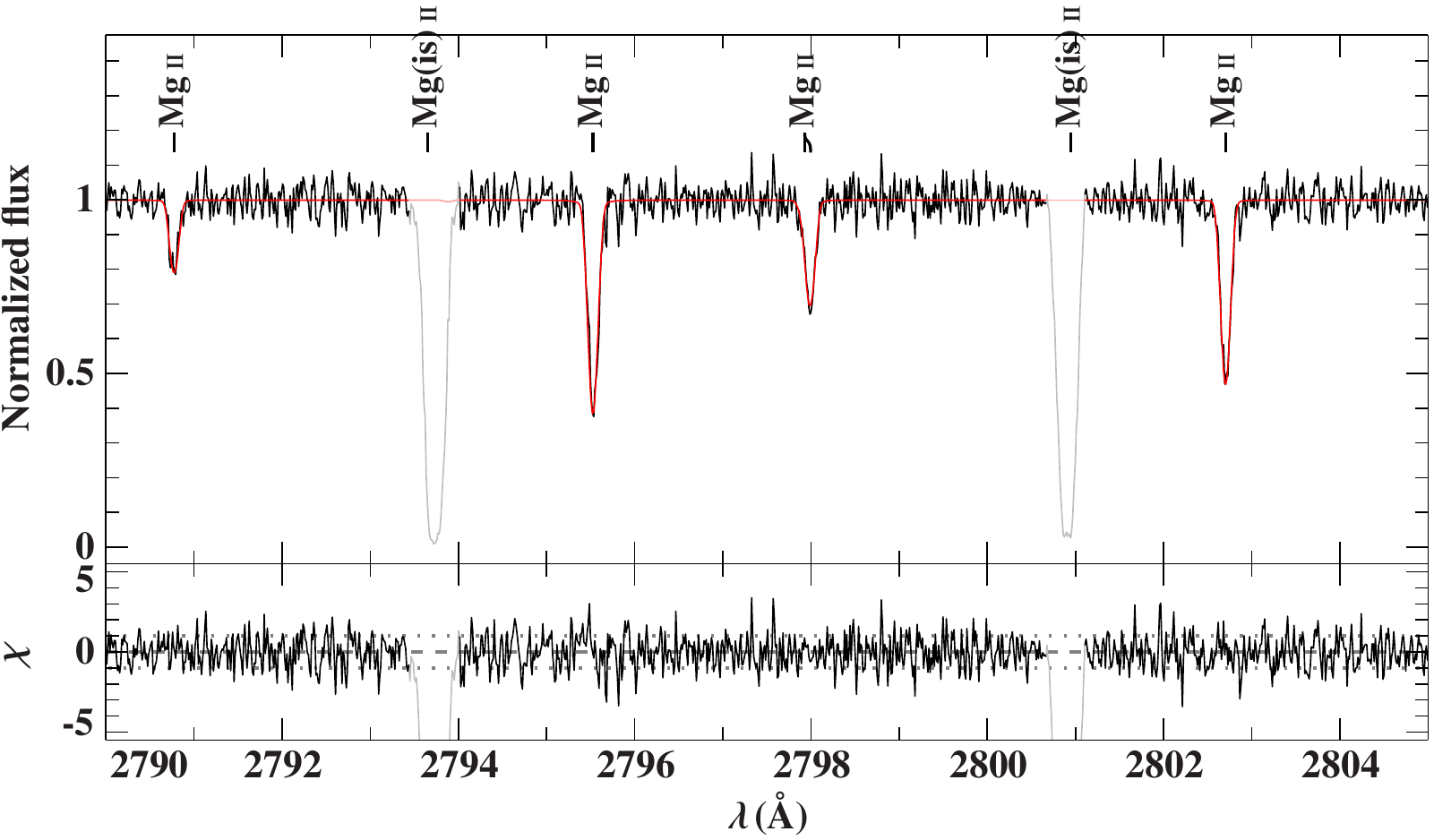}
\caption{Fit of the Mg \textsc{ii} lines in the NUV1 spectrum for $v_{\rm rot}$ sin $i$ = 7.9 km s$^{-1}$ and $v_{\rm micro}$ = 1.1 km s$^{-1}$. The IS Mg lines are shown in grey and were not included in the fit.}
\label{mgvsini}
\end{figure*}

\section{Spectroscopic analysis}

\subsection{Model atmospheres and synthetic spectra}

The model atomspheres used for our analysis are computed following the hybrid non-LTE (NLTE) approach discussed in \citet{przy06} and \citet{nieva07,nieva08}; it is a  combination of three codes: \textsc{Atlas}, \textsc{Detail}, and \textsc{Surface} (hereafter  ADS). The structure of the atmosphere, such as the temperature stratification and density, is based on line-blanketed, plane-parallel, homogeneous, and hydrostatic LTE-model atmospheres computed with the \textsc{Atlas12} code \citep{kur96}. The NLTE effects are accounted for in the computation of the atomic population numbers via updated versions of the \textsc{Detail} and \textsc{Surface} codes \citep{gid81}.
 Using the \textsc{Atlas12} LTE atmospheric structure as input, \textsc{Detail} solves the coupled radiative transfer and statistical equilibrium equations to obtain population numbers in non-LTE. This is then used by \textsc{Surface}, which produces the final synthetic spectrum using more detailed line-broadening data. 
%Though never used before for analysis of white dwarfs, 
The ADS approach is successfully used to study some types of hot stars (e.g., OB-type stars and BA-type supergiants; \citealt{nieva12}). It has also been tested in the sdB regime \citep{przy06}. Compared to the fully NLTE approach, the hybrid approach is much less time-consuming in terms of computation. Given the atmospheric parameters of \hd, no strong NLTE effects are expected on the temperature--density stratification of the atmosphere and the ADS models are appropriate. It must be kept in mind that an NLTE computation of atomic populations requires specific model atoms. Such model atoms are available only for a limited number of atomic species (see Sect. 4.3). When no such models exist, the populations are computed assuming the LTE approximation.

Given the very low metallicity expected for the star, a metallicity of 3\% (by numbers) solar was used in the computation of the \textsc{Atlas12} LTE model atmospheres. Afterward, small grids of different abundances were computed for all the atomic species seen in our spectra. It should also be mentioned that the He abundance was kept fixed at the value of log \nhe\ = $-$5.0 derived in \citet{heb03}. 
In order to generate spectra including the lines of all visible elements, normalized synthetic spectra computed for the various elements were multiplied following the technique described in \citet{irr14}. This approach is appropriate when blending of lines originating from different species is sparse, as  is the case for \hd. 
The metallic lines are then fitted using a standard $\chi^2$ minimization technique.

\subsection{Rotational velocity measurement}

As a first step, we measured the rotational velocity ($v_{\rm rot}$ sin $i$) using the Mg \textsc{ii} lines in the NUV spectra, as well as the Fe lines in the FUV spectra. At the same time, we also fitted the microturbulent velocity as it can contribute to line broadening, even though it is not expected to be very important for hot subdwarf stars because of their stable and radiative atomspheres.
Microturbulence has to be taken into account during the model atmosphere computation, as opposed to the rotational broadening that is applied afterward to the synthetic spectra. So we computed models with microturbulence between 0 and 8 km s$^{-1}$ only for Mg and Fe. Microturbulence affects the strong and weak lines in  different ways, so the choice of Mg and Fe allowed  lines of various strength to be 
fit in the NUV and FUV spectra.
% , so in addition to Mg, we chose iron to help constraining $v_{\rm micro}$, as it shows lines of various strength in the FUV spectrum. , and was varied between 0 and 8 km s$^{-1}$. 
The rotational broadening, microturbulence, magnesium, and iron abundances were thus fitted simultaneously. The microturbulence could be constrained to $v_{\rm micro}$ $\la$ 2 km s$^{-1}$; its effect below this value is rather small and did not influence  the resulting $\chi^{2}$ much.
We measured a rotational broadening of $v_{\rm rot}$ sin $i$ = 7.9 $\pm$ 0.3 km s$^{-1}$. The resulting fit for the Mg lines in the NUV1 spectrum and the residual of the fit are shown in Fig. \ref{mgvsini}. The IS lines are plotted in grey and were  excluded from the fitting procedure.

If it is assumed that the system is in synchronous rotation, then the star's rotation period is the same as the orbital period. Using the mass of the star to determine the radius, the equatorial rotation velocity can be derived: 
\begin{equation}
v_{\rm rot}=\frac{2 \pi R}{P}.
\end{equation}

Combined with the measured $v$ sin $i$, it yields the inclination angle. For a mass of 0.21 \msun, as predicted by theoretical models, we find $i$ = 57$\degr$ $\pm$ 4$\degr$ and a companion mass of 1.0$^{+0.10}_{-0.08}$ \msun. 

If we use the slightly higher sdB mass resulting from the Hipparcos parallax (0.245 \msun), we find $i$ = 52$\degr$ $\pm$ 3$\degr$ and a companion mass of 1.2$^{+0.13}_{-0.09}$ \msun \footnote{Only the uncertainty on $i$ is considered in these estimates of the companion mass.} .

\subsection{Metal abundance determinations}

Our FUV spectrum reveals a  number of lines from various elements, namely Si, Al, P, S, Ti, Cr, Mn, Fe, Ni, Zn, Ga, Sn, and Pb. The NUV spectrum is rather poor in metallic lines: besides the strong Mg lines, only a handful of weak Fe~\textsc{ii} lines are visible. 
Our model atmospheres were computed using the parameters derived by \citet{heb03}: 
\teff\ = 21,500 $\pm$ 500 K, log $g$ = 5.66 $\pm$ 0.06, and log \nhe\ = $-$5.0. Our grid also included the extreme parameters resulting from the uncertainties on \teff\ and log $g$, which were used to check the effects of changing those parameters on the metallic abundances. Microturbulence was kept to zero in our metal grids and $v$ sin $i$ = 7.9 km s$^{-1}$ was used during the fitting procedures.

As mentioned earlier, in order to compute NLTE population numbers, detailed model atoms are needed. To date, the model atoms available for \textsc{Detail}  cover C, N, O, Mg, Al, Si, S, and Fe, so these elements were treated in NLTE in \textsc{Detail}, while all others were computed using the LTE approximation. 
We note that the LTE approximation is expected to be adequate given the atmospheric parameters of the star \citep{przy06b} and it is the usual approach for abundance analysis of white dwarfs and hot subdwarfs having similar or higher \teff. 
Nevertheless, we also computed LTE models for Mg, Al, Si, S, and Fe in order to compare abundances derived with both types of models. This is discussed in more detail in Sect. 5.1. 
Our resulting abundances are listed in Table \ref{abund} where the results obtained with NLTE and LTE models are indicated in two separate columns. The comparison between our final synthetic spectrum and the entire UV spectrum can be found in Fig.~\ref{spectra}.

\begin{table}[b]
\caption{Abundances of metals detected in the spectra of \hd.} \label{abund}
\centering
%\scriptsize
\begin{tabular}{l c c}
\hline
\hline
Element & \multicolumn{2}{c}{Abundance} \\
Z & \multicolumn{2}{c}{log $N$(Z)/$N$(H)} \\
        &  NLTE   & LTE \\
\hline 
C~\textsc{ii}   & \textless $-$9.6 &  $-$\\
N~\textsc{ii}\tablefootmark{a}  & \textless $-$5.0 &  $-$ \\
O~\textsc{i}    & \textless $-$6.3 & $-$\\
Mg~\textsc{ii}  & $-$6.40 $\pm$ 0.07 & $-$6.20 $\pm$ 0.07 \\
Al~\textsc{iii} & $-$7.37 $\pm$ 0.05 & $-$7.39 $\pm$ 0.05 \\
Si~\textsc{ii-iii-iv}   & $-$7.3 $\pm$ 0.1 & $-$7.6 $\pm$ 0.1 \\
P~\textsc{iii}  & $-$ & $-$9.2 $\pm$ 0.1 \\
S~\textsc{ii}   & $-$8.0 $\pm$ 0.1 & $-$8.4 $\pm$ 0.1 \\
Ca~\textsc{ii}\tablefootmark{a} & $-$7.3 $\pm$ 0.2 & $-$ \\
Ti~\textsc{iii} & $-$ & $-$8.30 $\pm$ 0.05 \\
Cr~\textsc{iii} & $-$   & $-$7.55 $\pm$ 0.05 \\
Mn~\textsc{iii} & $-$ & $-$8.00 $\pm$ 0.05 \\
Fe~\textsc{ii-iii} & $-$5.75 $\pm$ 0.15 & $-$6.1 $\pm$ 0.1\\
%Fe~\textsc{ii} (nuv)  & $-$5.63 $\pm$ 0.1 \\
%Fe~\textsc{iii} & $-$5.85 $\pm$ 0.1 \\
Ni~\textsc{iii} & $-$6.6 $\pm$ 0.1 & $-$6.7 $\pm$ 0.1  \\
Zn~\textsc{iii} & $-$ & $-$7.72 $\pm$ 0.07 \\
Ga~\textsc{ii-iii} & $-$        & $-$9.6 $\pm$ 0.1  \\
Sn~\textsc{iii} & $-$   & $-$10.6 $^{+0.6}_{-0.1}$  \\
Pb~\textsc{iv} & $-$    & $-$10.0 $\pm$ 0.2  \\
\hline
\end{tabular} \\
\tablefoottext{a}{Upper limit or abundance determined with optical spectra.}
\end{table}

\begin{figure*}[p]
\begin{center}
\includegraphics[width=16cm]{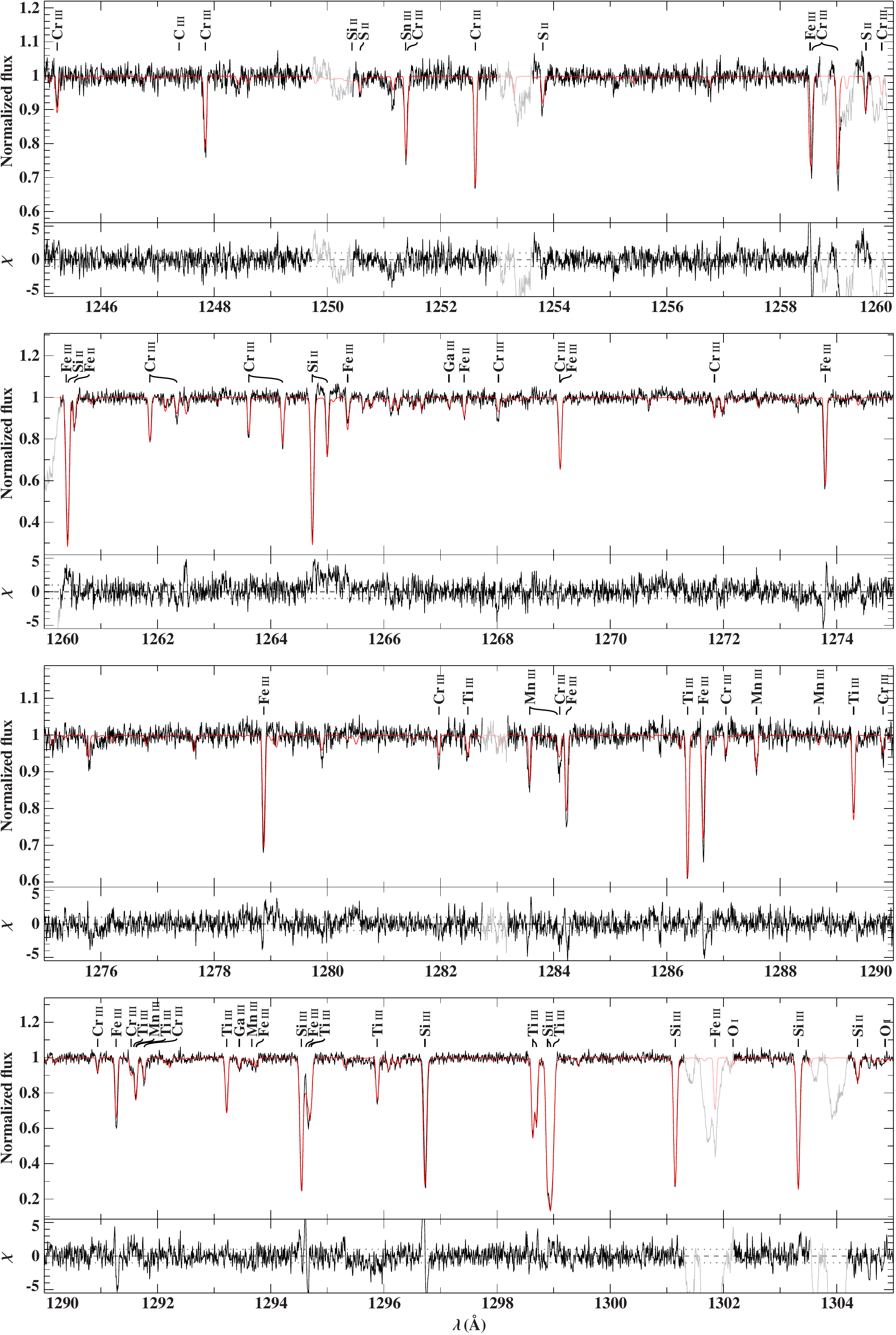}
\caption{Entirety of our UV spectra of \hd, compared with our best fitting model. 
The bottom of each panel features the difference between the observation and the model divided by the noise as $\chi$. 
The portions of the spectrum plotted in grey are regions excluded from the fits. 
Most of them feature blueshifted IS lines. The IS lines are artificially broadened by the RV correction applied to the individual spectra and show two components due to a time gap in the series of FUV spectra. For this panel, the abundance of Sn was set to the value matching the \ion{Sn}{iii} line. } 
\label{spectra}
\end{center}
\end{figure*}

\addtocounter{figure}{-1}
\begin{figure*}[p]
\begin{center}
\includegraphics[width=16cm]{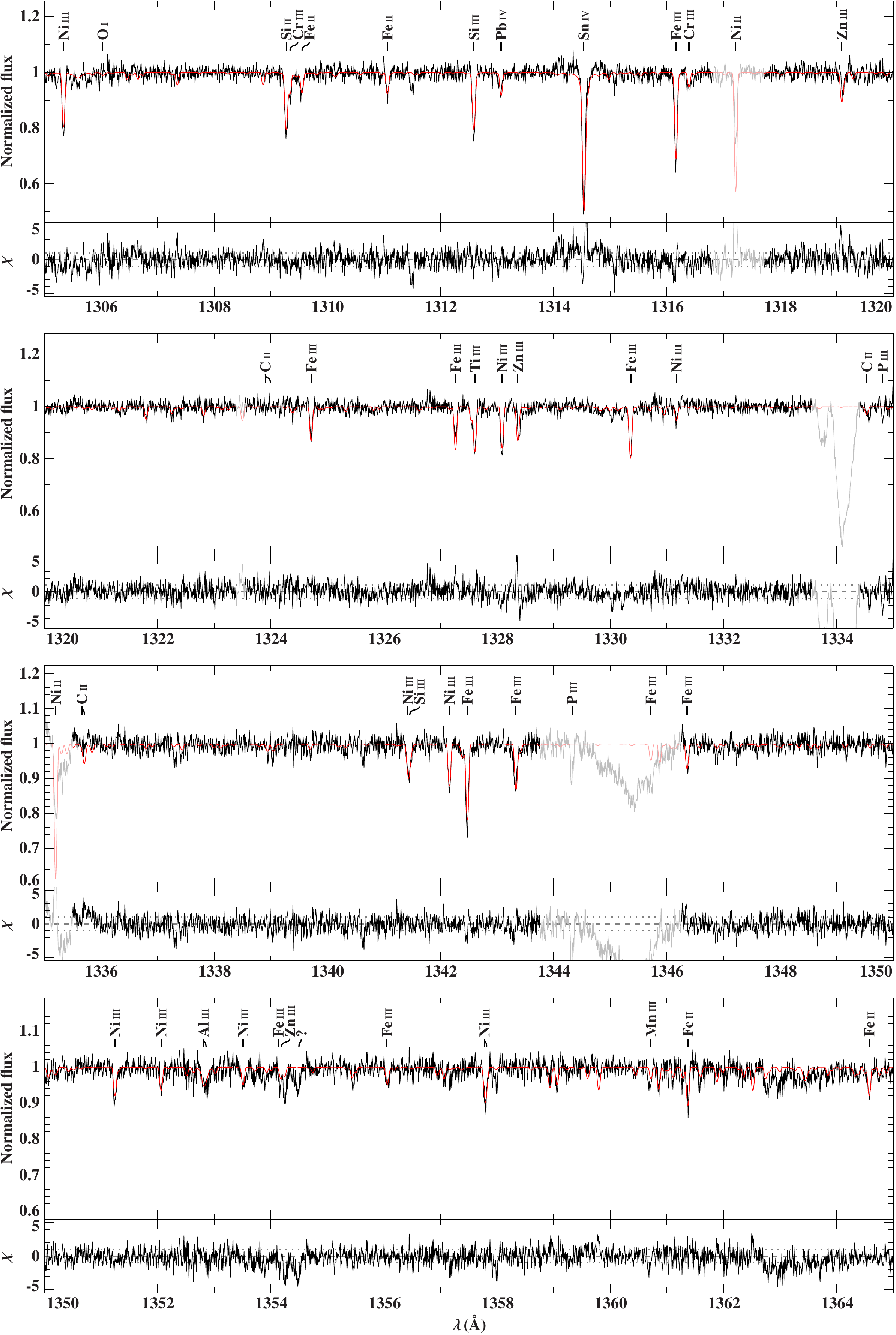}
\caption{Continued. Here and in the following panels, the Sn abundance is set to the value matching the \ion{Sn}{iv} lines. Nickel was included at the abundance determined by the \ion{Ni}{iii} lines, thus the \ion{Ni}{ii} lines appear too strong and were excluded from the fits (see Sect. 4.3.4). 
The broad feature at around 1345 \AA\ is an artifact (blemish) from the STIS detector.} 
\end{center}
\end{figure*}

\addtocounter{figure}{-1}
\begin{figure*}[p]
\begin{center}
\includegraphics[width=16cm]{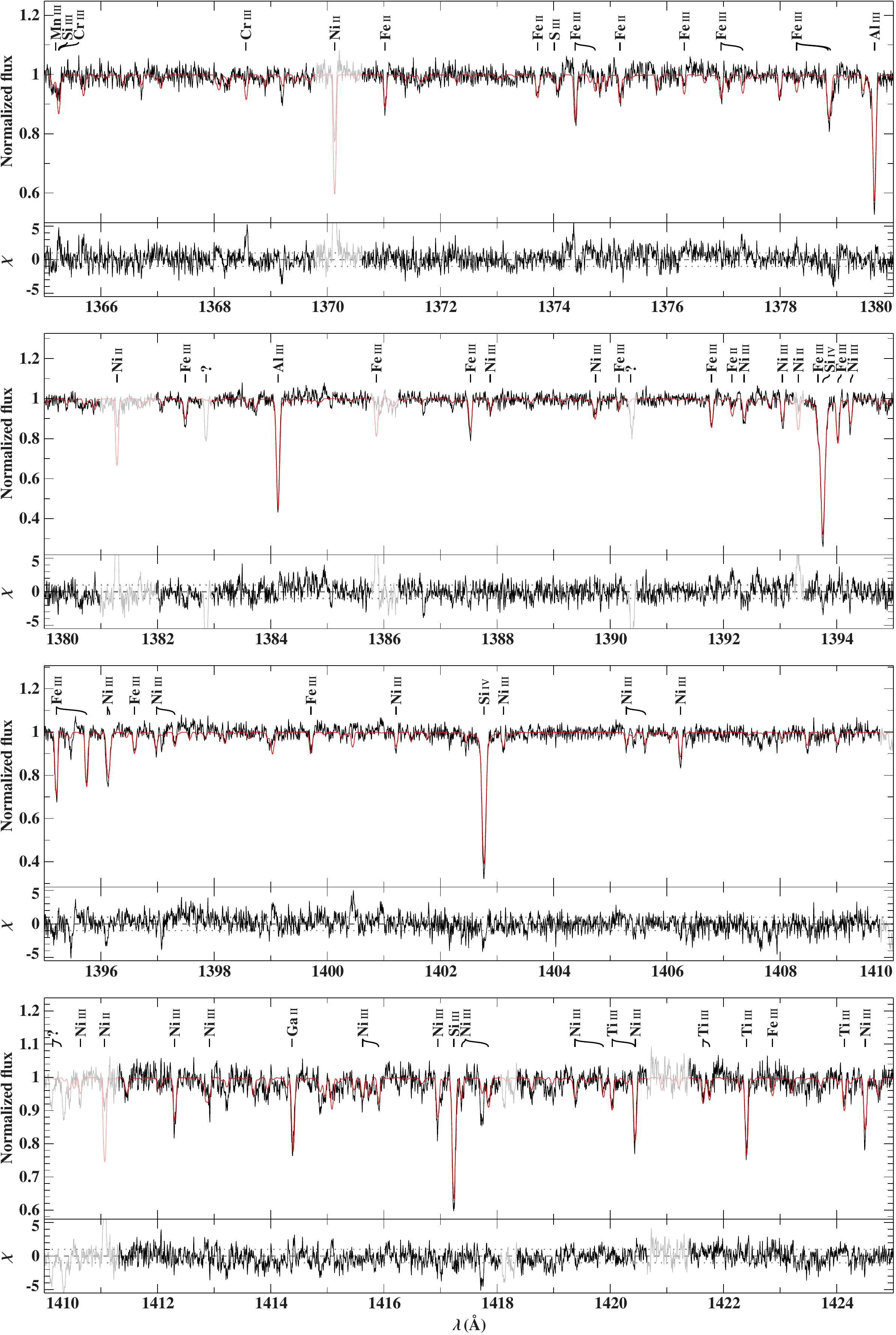}
\caption{Continued.} 
\end{center}
\end{figure*}

\addtocounter{figure}{-1}
\begin{figure*}[p]
\begin{center}
\includegraphics[width=16cm]{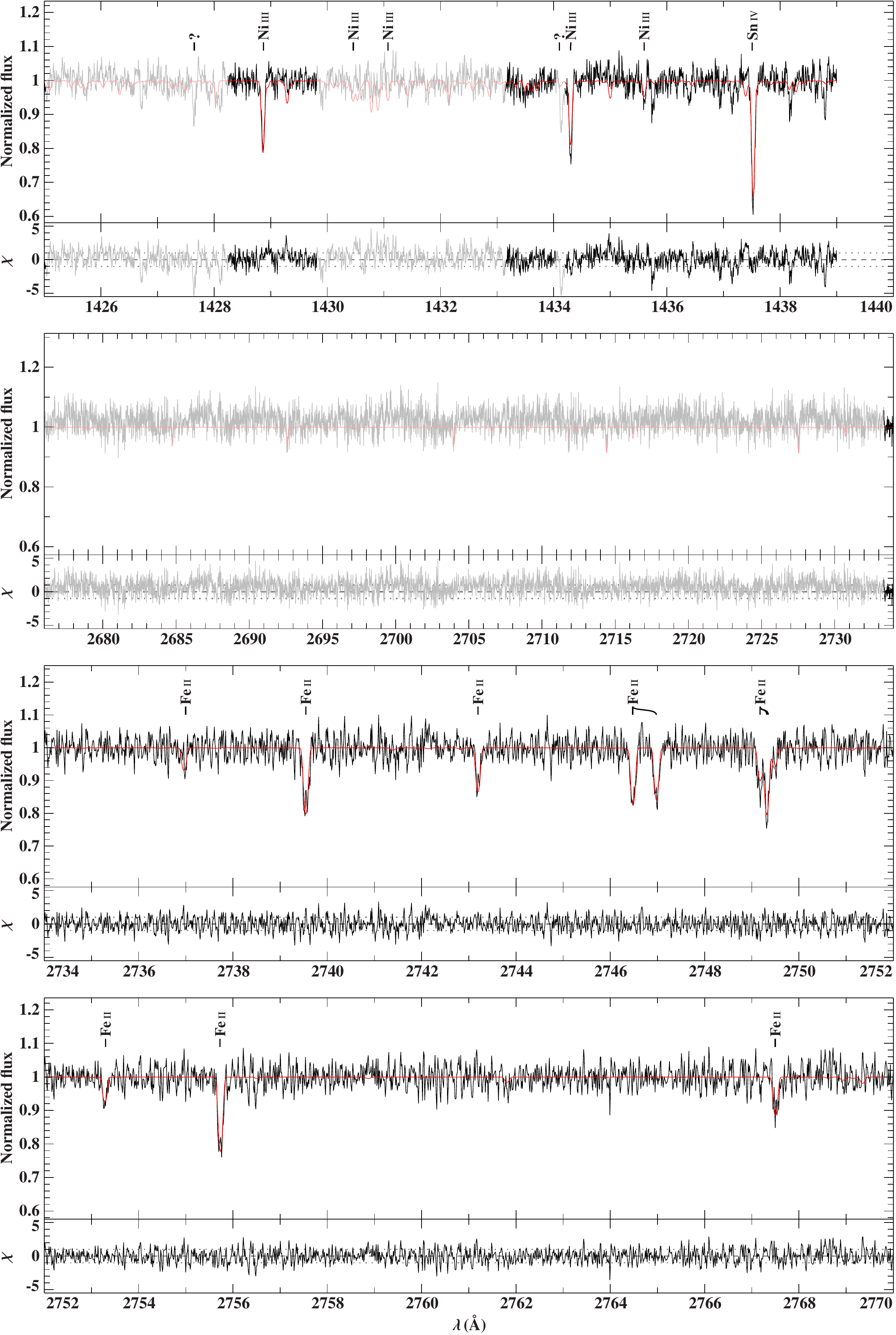}
\caption{Continued. Here starts the NUV2 spectrum. The scaling of the wavelength varies to include the spectral intervals devoid of fitted lines in unique panels. The iron abundance was set to the value derived with the \ion{Fe}{ii} lines.} 

\end{center}
\end{figure*}

\addtocounter{figure}{-1}
\begin{figure*}[t]
\begin{center}
\includegraphics[width=16cm]{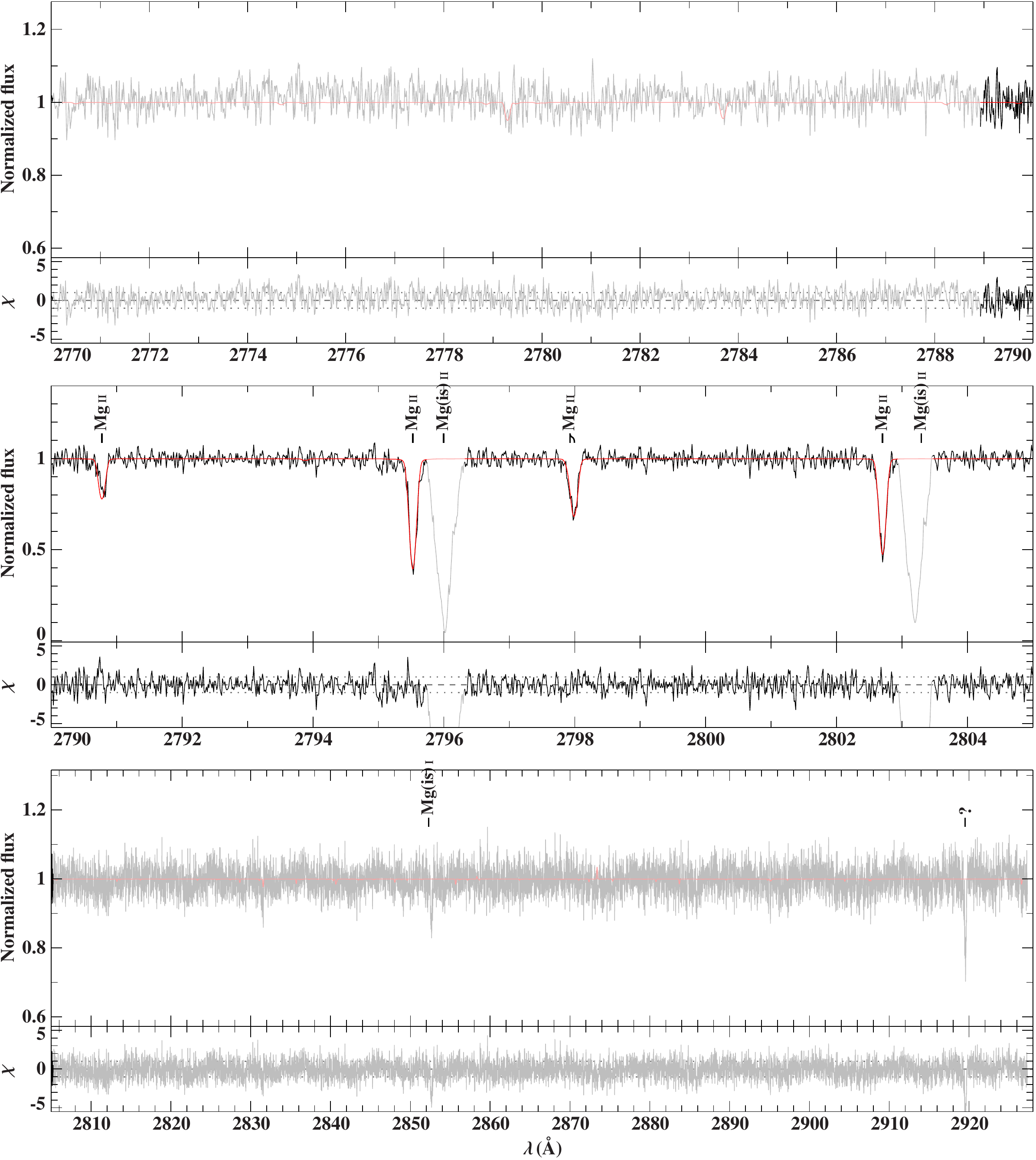}
\caption{Continued.} 

\end{center}
\end{figure*}

\subsubsection{Upper limits for C, N, and O}

No lines of the C, N, O elements are visible in the UV spectra. 
This is not very surprising in the cases of nitrogen and oxygen as no strong lines are predicted in our UV ranges. 

For oxygen, the strongest predicted features are three \ion{O}{i} lines at 1302, 1304, and 1305 \AA. They were used to put an upper limit of log $N$(O)/$N$(H) = $-$6.3. 
For nitrogen nothing useful is predicted in the UV range, so we turned toward the optical instead, where we used the high-resolution FEROS spectra of \hd\ obtained by \citet{edel05}. The \ion{N}{ii} lines at 3995 and 4630 \AA\ were used to put an upper limit of log $N$(N)/$N$(H) = $-$5.0, which is, however, only slightly below the solar value.

The situation is different for carbon. Three strong lines are predicted in the UV range, including \ion{C}{ii} resonance lines at 1334.5 and 1335.7 \AA. The absence of these lines in our spectrum results in an extremely low upper limit of log $N$(C)/$N$(H) = $-$9.6. This is six orders of magnitude below the solar value.
Carbon has essentially vanished from the surface of the star.

\subsubsection{Mg, Al, Si, P, and S}

\begin{figure*}[t]
\sidecaption
\includegraphics[width=12cm,clip=true]{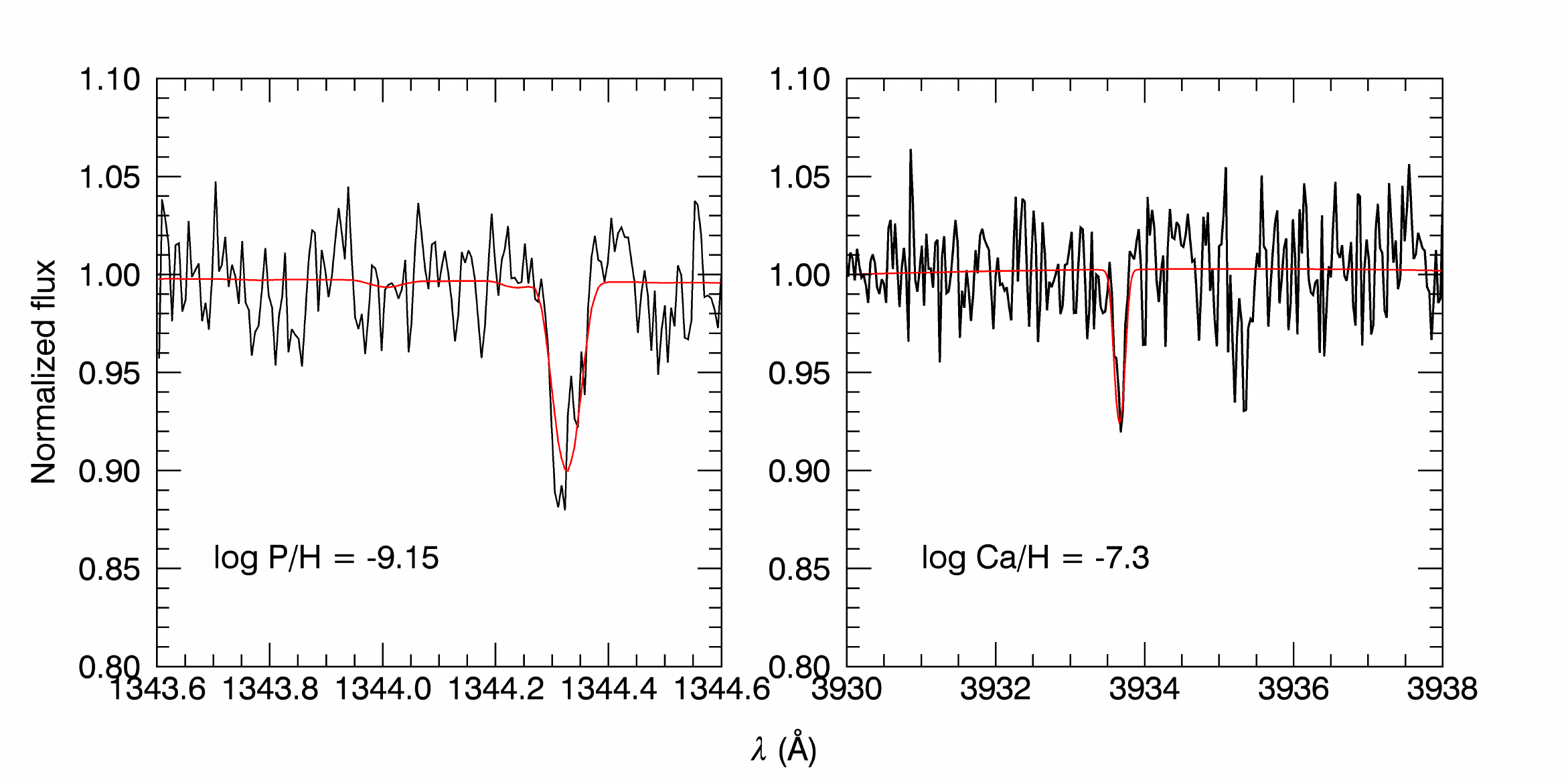}
%\resizebox{\hsize}{!}{\includegraphics{Pfitest.eps}}
\caption{Best fits for the \ion{P}{iii} line in the FUV spectrum as well as for the Ca~\textsc{ii k} line in an optical FEROS spectrum.} 
\label{fitp}
\end{figure*}

Only four Mg lines are visible in the NUV spectra. Their abundances were fitted during the determination of the projected rotational velocity (see Fig. \ref{mgvsini}). We also fitted them in the NUV2 spectra, shown in the last panel of Fig. \ref{spectra}. The fits with both spectra lead to very similar results, with an abundance of 1/100 solar. This is essentially the same abundance as was first estimated by \citet{heb03} based on the \ion{Mg}{ii} $\lambda$4481 line visible in the optical spectrum. 

The FUV spectrum shows the \ion{Al}{iii} doublet at 1379, 1384 \AA\ that matches an abundance of log $N$(Al)/$N$(H) = $-$7.37.  

Only two very weak sulphur lines are visible at the blue end of the FUV spectra, but they were nevertheless  used to constrain the abundance of this element to log $N$(S)/$N$(H) = $-$8.0. 

One phosphorus line was used to fit: \ion{P}{iii} $\lambda$1344.32. Because of the lack of a line list for this ion in \textsc{Surface}, we used the public codes TLUSTY and SYNSPEC\footnote{http://nova.astro.umd.edu/} to perform an independent fit of the line. 
TLUSTY allows the computation of NLTE stellar atmospheres, assuming plane-parallel geometry and hydrostatic and thermal equilibrium; SYNSPEC is then used to compute synthetic spectra, given an atmospheric structure \citep{lanz07}.
An NLTE model atmosphere was computed including hydrogen and helium only and as no model atom is available for \ion{P}{iii}, we included it in an LTE way through \textsc{Synspec}. The resulting fit can be seen in Fig. \ref{fitp}. 

Silicon is an interesting element since it shows strong lines originating from three different ionization stages. Lines originating from all three of them are well reproduced with an abundance of log $N$(Si)/$N$(H) = $-$7.3. This is a good sign that the ionization equilibrium for this element is correct given the hybrid NLTE approach and the effective temperature of the star. Interstellar components to \ion{Si}{iii} $\lambda\lambda$1301, 1303 are present, but are not blended with the photospheric lines due to the relatively high radial velocity of the star during the FUV exposures. The IS features seen in our spectra are broadened by the radial velocity correction applied to the individual exposures, which explains their unusual shape in Fig. \ref{spectra}.

\subsubsection{Ca}

Calcium is an important element because it has been claimed that all ELM WDs with log $g$ $\la$ 5.9 show Ca in their optical spectra \citep{her14,gia14}. This was first illustrated by \citet{kap13} who combined results from different studies of ELM WDs in their figure 6. However, they erroneously indicated that \hd\  showed \ion{Ca}{ii} in its optical spectra, confusing it with the \ion{Mg}{ii} feature mentioned in \citet{heb03}. Nevertheless, since the tendency is true for all other ELM WDs analyzed so far, we went back to the optical spectra of \hd\ to look for this expected Ca~\textsc{ii} K line. After correcting a few spectra for their radial velocity, and scrupulously comparing them, we indeed found weak calcium features: one stable line at the predicted wavelength, thus of photospheric origin, and one moving around, being a very small IS component. As can be seen in Fig. \ref{fitp}, this line is  very weak indeed, but present in the different spectra, thus real. It is understandable why it was first overlooked in \citet{heb03}. It must also be kept in mind that most of the ELM WDs in which Ca has been measured have \teff\ \textless\ 16,000 K. At high temperatures, the strength of these lines rapidly 
decreases as \ion{Ca}{ii} becomes ionized. Nevertheless, the tiny line we found in \hd\ indicates an abundance of log $N$(Ca)/$N$(H) = $-$7.3. 
Calcium was also measured in another ELM WD (SDSS J0815+2309) having essentially the same temperature as \hd\ but a slightly higher surface gravity  \citep{gia14}. Compared to \hd, the Ca abundance derived for J0815+2309 is a hundred times higher (log $N$(Ca)/$N$(H) = $-$5.27). As for other ELM WDs, the Ca abundance of \hd\ is similar to the lowest abundances measured in the sample of \citet{gia14}.

\subsubsection{The iron-group elements}

Titanium, chromium, iron, and nickel show many strong lines from their main ionization stage (\textsc{iii}) that could be  reproduced well by our synthetic spectra. Two well-defined lines of \ion{Mn}{iii} are seen at 1283 and 1287 \AA, and two lines of \ion{Zn}{iii} are also visible at 1318 and 1328 \AA. The resulting abundances are listed in Table \ref{abund}.

Iron and nickel are the only iron-group elements that also show lines from their singly ionized stage. There is a small discrepancy in the abundances measured with the \ion{Fe}{iii} lines in the FUV and the \ion{Fe}{ii} lines mainly found in the NUV. Compared with the abundance of log $N$(\ion{Fe}{iii})/$N$(H)\footnote{We use the nomenclature \ion{Fe}{iii} here and below to indicate the abundance derived by using lines from a specific ion.} = $-$5.87 $\pm$ 0.07, the \ion{Fe}{ii} lines need a slightly \textit{higher} abundance of $-$5.6 $\pm$ 0.1 to be reproduced. 
Another investigation of UV iron lines in B-type stars using \textsc{ADS} models
found a scatter of $\sim$0.25 dex for FUV lines when abundances measured from different lines are compared \citep{schaf15}. The reason for this, possibly related to the \textsc{ADS} models, will be explored further in a future publication (Schaffenroth et al. 2016, in preparation).
In any case, this scatter is similar to the discrepancy we found between the FUV and NUV lines of \hd. We thus adopt a mean abundance of $-$5.75 $\pm$ 0.15 for iron, as indicated in Table \ref{abund}.

We also encountered a discrepancy between lines originating from both ionization stages of nickel, but this time with a significant difference. In the case of nickel, fitting the \ion{Ni}{ii} lines results in an abundance of log $N$(\ion{Ni}{ii})/$N$(H) = $-$7.25, which is 0.55 dex \textit{below} what is predicted by the main ionization stage (log $N$(\ion{Ni}{ii})/$N$(H) = $-$6.7 when using the \textsc{ADS} models). In other words, at the abundance derived from the \ion{Ni}{iii} lines, the \ion{Ni}{ii} lines are much  too strong, as can be seen in Fig. \ref{spectra}. Something to keep in mind is that, unlike iron, nickel is treated using the LTE approximation in \textsc{ADS}. 
As nickel model atoms are available for TLUSTY (but only for \ion{Ni}{iii} to \ion{Ni}{vi}), we built a small grid of NLTE models with various Ni abundances in order to fit the \ion{Ni}{iii} lines. We obtained an abundance of log $N$(\ion{Ni}{iii})/$N$(H) = $-$6.6 $\pm$ 0.1 dex, essentially the same as derived with \textsc{ADS}. This is not surprising since \ion{Ni}{iii} is the main ionization stage. It is also reassuring in the sense that we can expect the LTE abundances derived for the other iron-group elements via their main ionization stage to be reliable.
However, owing to the lack of a \ion{Ni}{ii} model atom for TLUSTY, this ion was treated in LTE and an abundance discrepancy similar to the one seen with the \textsc{ADS} models was observed.

\subsubsection{Trans-iron elements: Ga, Sn, and Pb}

We found lines originating from three trans-iron elements, gallium, tin, and lead, in the spectra of \hd.
Such heavy elements are also seen in hot subdwarfs \citep{otoole04,otoole06} as well as hot DA and DO WDs \citep{ven05,wer12,rauch13}. However, they have not been detected yet in colder WDs, with the exception of the heavily polluted WD GD 362, in which a strontium line has been detected \citep{zuck07}. The metallic elements in this cool star are accreted from a debris disk surrounding the star. With its temperature of 21,500 K, \hd\ is among the coolest WD/sdB stars for which trans-iron elements have been detected. 

The Ga abundance of log $N$(Ga)/$N$(H) = $-$9.55 was determined using the $\lambda$1414.4 resonance line of \ion{Ga}{ii} as well as two weak \ion{Ga}{iii} lines at 1267 \AA\ and 1293 \AA. All of them are  reproduced well with this abundance.

A sole line of lead is present in our spectra, the 1313 \AA\ resonance line of \ion{Pb}{iv}, which indicates an abundance of log $N$(Pb)/$N$(H) = $-$10.0, slightly above the solar value.
%\citet{otoole04} looked for heavy elements in the UV spectra of 27 hot subdwarf stars and could identify lead lines in 22 of them as for the 5 remaining stars, the presence of lead remained uncertain. Nevertheless this study pointed out that this element is commonly seen in such stars. Abundances were measured in a dozen sdBs and appears to be enriched in all of them.   

We see the resonance line of \ion{Sn}{iii} at 1251.39 \AA, as well as the resonance doublet of \ion{Sn}{iv} $\lambda\lambda$ 1314, 1437. 
The fit of the \ion{Sn}{iii} line gives an abundance of $N$(Sn)/$N$(H) = $-$10.6.
However, at this abundance, the \ion{Sn}{iv} lines are barely visible in our synthetic spectra. In order to reproduce them, we need an abundance of $-$8.4, which is more than 2 dex higher than what is indicated by the \ion{Sn}{iii} line. This is reminiscent of the problem encountered with nickel, but much more dramatic.
Given the fact that these three transitions are resonances, their oscillator strengths, taken from \citet{mor00} are considered to be reliable. The \ion{Sn}{iii} line is blended with a \ion{Cr}{iii} line (1251.417 \AA), which is however predicted to be rather weak at the Cr abundance determined. Indeed, a fit of the \ion{Sn}{iii} line without the contribution of the \ion{Cr}{iii} feature leads to an abundance only slightly higher, $-$10.5 dex, still far from the $-$8.4 dex needed to reproduce the \ion{Sn}{iv} lines.
As we did for nickel, we turned toward NLTE models in an attempt to solve the tin abundance discrepancy.
An NLTE fit of the Sn lines was performed using this time the T\"{u}bingen NLTE model-atmosphere package (TMAP; \citealt{tmap03}) because a simple NLTE Sn model atom for this code was built in the course of the spectral analysis of the hot DA white dwarf G191-B2B (\citealt{rauch13}; T. Rauch, private communication). With NLTE models, the \ion{Sn}{iii} line leads to an abundance of $-$10.0, 
%which is higher than the LTE abundance of $-$10.6, 
and the \ion{Sn}{iv} lines indicate $-$8.2 (1314.5 \AA) and $-$8.5 (1437.5 \AA). %which is essentially equivalent to what we derived with \textsc{ADS}. 
According to these results, NLTE effects apparently cannot solve the ``tin problem''. However, because the model atom for tin is rather simple (3 and 6 NLTE levels for \ion{Sn}{iii} and \textsc{iv}, respectively) and resonance lines can be very sensitive to NLTE effects, it would be premature to exclude these effects as the cause of the discrepancy and to blame it on the atomic data. 
Because \ion{Sn}{iii} has the same ionization energy as \ion{Fe}{iii} and \ion{Cr}{iii}, it is expected to be the main ionization stage, as for the iron-group elements, so we adopt an abundance of log $N$(Sn)/$N$(H) = $-$10.6$^{+0.6}_{-0.1}$. This is based on our ADS results, but the upper limit includes the value indicated by TMAP.  

\section{Discussion}

\begin{figure*}
\begin{center}
\includegraphics[width=16cm]{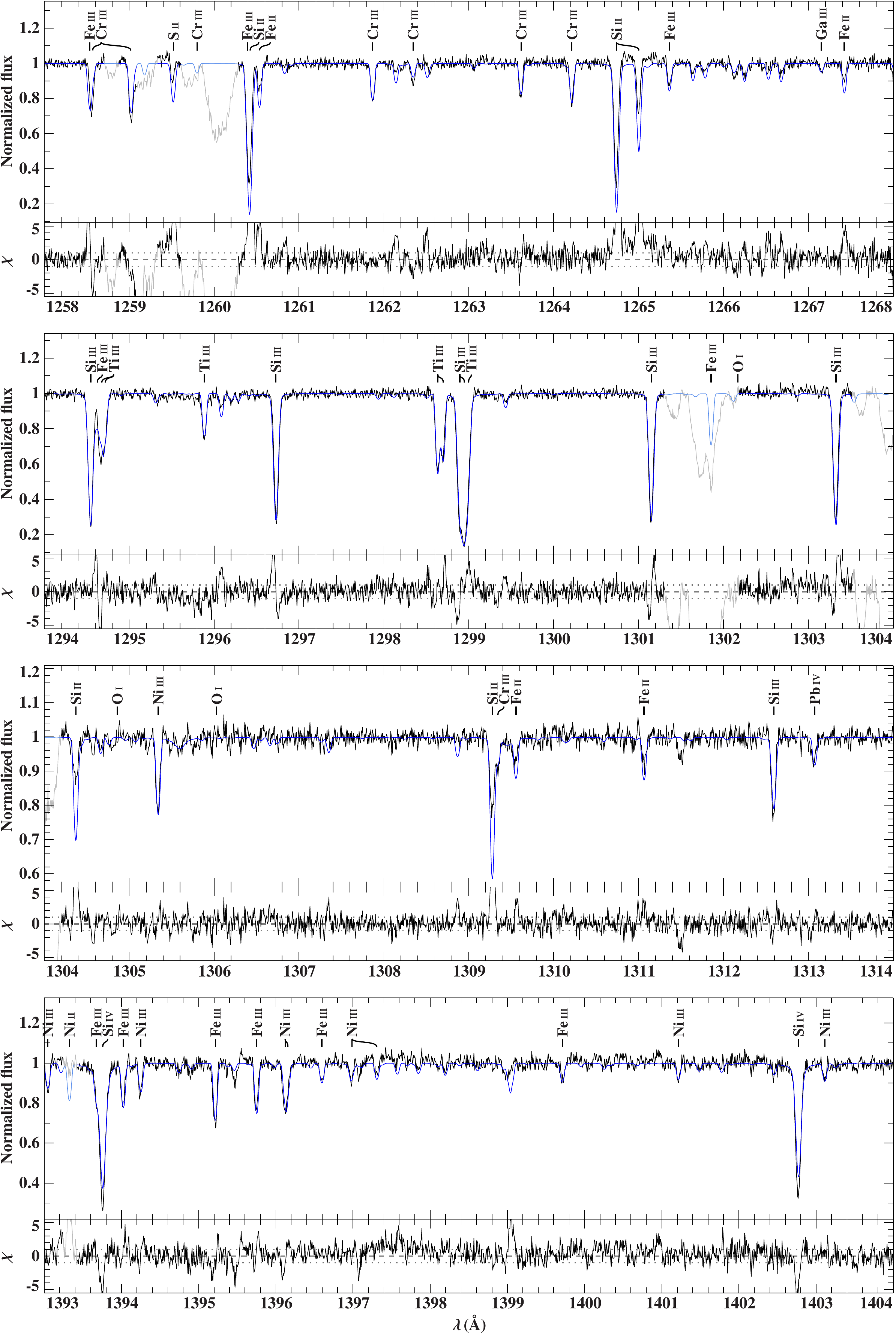}
\caption {Comparison between parts of the FUV spectrum featuring mainly the Si lines and an LTE model spectrum, in blue. The LTE model was computed using the NLTE abundances stated in Table \ref{abund}. The \ion{Si}{ii}, \ion{Fe}{ii}, and \ion{S}{ii} lines appear to be much stronger than their observed counterparts, while the \ion{Si}{iii} lines are nicely reproduced.}
\label{lte}
\end{center}
\end{figure*} 

\begin{figure*}[th]
\begin{center}
\includegraphics[width=16cm]{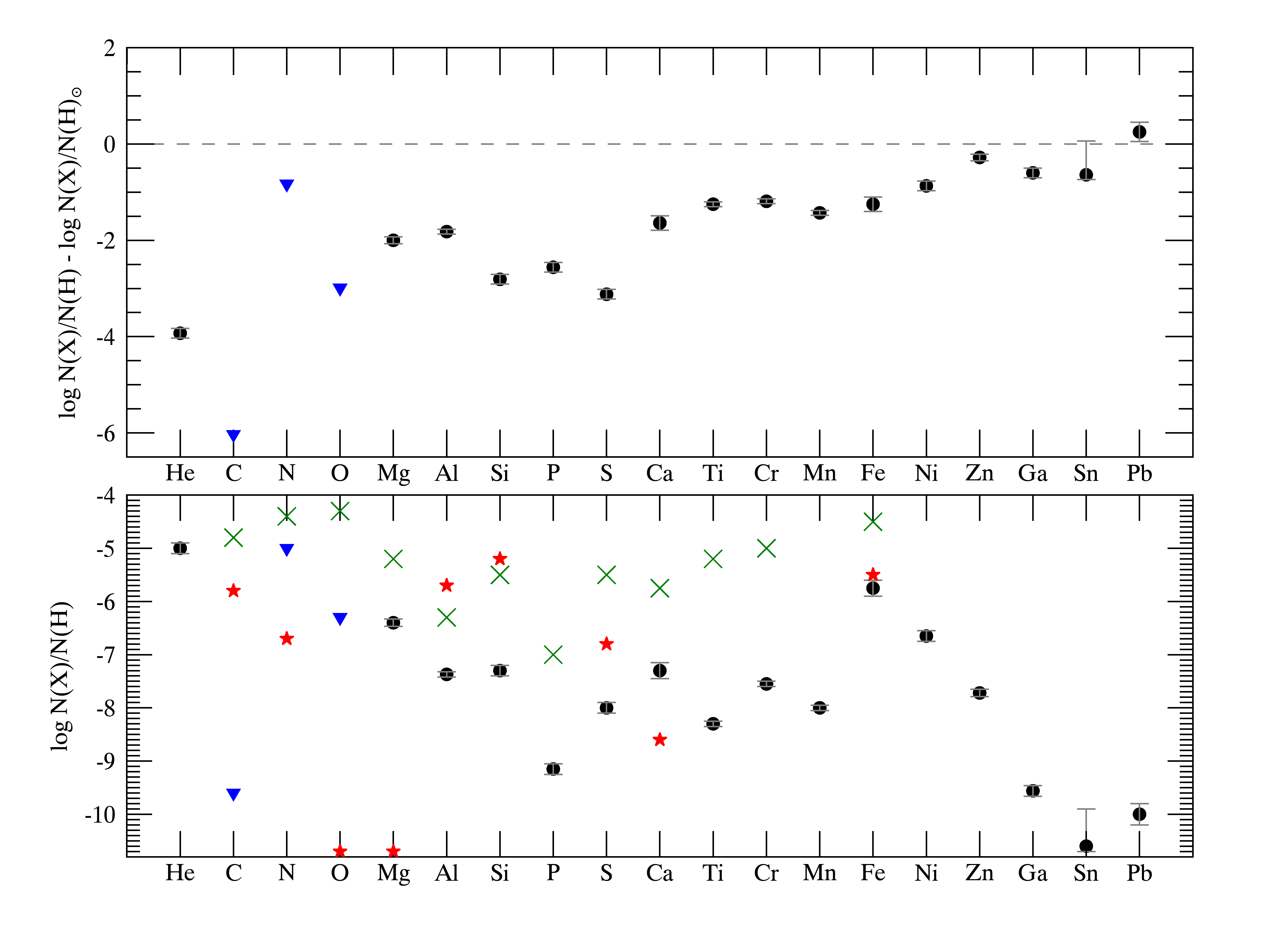}
\caption{Summary of the determined chemical composition of \hd. The top panel shows the abundances relative to the solar ones \citep{asp09} and the bottom panel the absolute values. Down-pointing triangles indicate upper limits determined for C, N, and O. The stars in the bottom panel indicate abundances predicted by radiative levitation (P. Chayer, private communication). No support is expected for oxygen and magnesium, thus the stars are at the bottom of the plot. The green crosses indicate the average abundances for sdB stars.} 
\label{comp}
\end{center}
\end{figure*}

\subsection{Non-LTE effects}

The discrepancies found between the ionization stages of Ni and Sn show the same trend: the lowest ionization stage visible indicates a lower abundance. This  is sometimes seen in much hotter stars and is a signature of NLTE effects when analyses are made using the LTE approximation. As stated in \citet{dre93}, in a consistent NLTE model, ionization equilibrium is shifted to higher stages. In our case, this could explain why the \ion{Ni}{ii} lines are too strong in our models, while the \ion{Sn}{iv} lines are too weak.
However, these effects are not really expected to appear in an sdB at 21,500 K, although some line-by-line differences between LTE and NLTE treatment were found in HD 205805, an sdB star at 25,000 K and log $g$ = 5.0 \citep{przy06b}. 
This is why we decided to compute additional LTE models for Mg, Al, Si, S, and Fe and to redo the fitting procedure in order to derive LTE abundances for these elements\footnote{We remind the reader that the hybrid approach of ADS only includes NLTE effects in the computation of the population numbers; in all cases (NLTE and LTE) the atmospheric structure is computed assuming the LTE approximation.}.
The resulting abundances are indicated in Table \ref{abund}. For magnesium and aluminium, the LTE and NLTE abundances are  similar, while there is a difference of 0.4 dex for the sulfur abundance. Most interesting is silicon;
a simultaneous fit of all the Si lines leads to an abundance of log $N$(Si)/$N$(H) = $-$7.56 dex, but the quality of the fit is poor and it shows conspicuous hints of an incorrect ionization equilibrium. Instead, the lines originating from the different ions had to be fitted separately in order to be correctly reproduced, thus leading to discrepant abundances:\\
  log $N$(\ion{Si}{ii})/$N$(H) = $-$7.8 $\pm$ 0.1, \\
  log $N$(\ion{Si}{iii})/$N$(H) = $-$7.33 $\pm$ 0.05, \\
 and log $N$(\ion{Si}{iv})/$N$(H) = $-$7.1 $\pm$ 0.1. \\
The abundance of the main ionization stage (\ion{Si}{iii}) remains the same as derived in NLTE, but the \ion{Si}{ii} abundance is 0.5 dex lower. 
The \ion{Fe}{ii} lines in the NUV spectrum also lead to an LTE abundance 0.5 dex lower than the value derived with NLTE models. 
To illustrate this effect on the spectral lines, we select four spectral ranges featuring Si lines and show the corresponding LTE spectrum in Fig. \ref{lte}. The spectrum includes the abundances determined with the NLTE models. The \ion{Si}{ii} lines (as well as few \ion{Fe}{ii} lines and one \ion{S}{ii} line) appear too strong, as can be seen with the \ion{Ni}{ii} lines (Fig \ref{spectra}). 

It thus becomes clear that NLTE effects are present in \hd\ and influence the line strengths, especially those of the non-dominant ionic species. For \ion{Si}{ii}, \ion{S}{ii}, and \ion{Fe}{ii} the differences in strength of the modeled lines between LTE and NLTE lead to a $\sim$0.5 dex shift in abundances.
However, the LTE abundances derived for the main ionization stage (in this case doubly ionized) are consistent with the NLTE values. This result confirms that the abundances derived for the elements that could not be treated in NLTE can be regarded as reliable since the main ionization stage was used. The only exception is lead, for which only one \ion{Pb}{iv} line was visible. In this case the derived abundance might be overestimated.
The strongest gallium line is from \ion{Ga}{ii,} but the LTE models simultaneously reproduced this line and the two weak \ion{Ga}{iii} features.
However, some mystery still remains about the Sn lines: it is not clear what is causing the 2 dex discrepancy between \ion{Sn}{iii} and \ion{Sn}{iv}.
 This ``tin issue'' is certainly worth further investigation in the future. Maybe more sophisticated NTLE model atoms (including more energy levels) would solve the problem, or perhaps the collisional radiative rates used for tin are not very reliable. There is also the possibility that the atomic data are inaccurate.

\subsection{Metal abundances}

Figure \ref{comp} summarizes the chemical composition of \hd\  with upper limits for the C, N, O elements, and abundance values for 15 metallic elements. All of the elements, except lead, are found to have subsolar abundances, with the heavy elements (Ti and heavier) being the most abundant relative to solar (see top panel of Fig. \ref{comp}). Lead is, however, slightly above the solar value. This  abundance pattern suggests that diffusion mechanisms counteract gravitational settling in order to keep a relatively considerable amount of heavy elements in the photosphere.

\subsubsection{Comparison with diffusion models}
The bottom panel of Fig. \ref{comp} also features abundances predicted by radiative levitation (red stars) for the atomic species included in TOPbase\footnote{http://cdsweb.u-strasbg.fr/topbase/topbase.html}. These abundances were computed for a model with \teff\ = 21,000 K and log $g$ = 5.6 according to the method described in \citet{cha95} (P. Chayer, private communication). In most cases (C, Al, Si, S, Fe) we measured abundances below the value predicted by radiative support, though the Fe abundance is quite close to the prediction. The two exceptions are Mg and Ca. Calcium is  1.5 dex higher than the predicted abundance while no support at all is expected for magnesium. 

\subsubsection{Comparison with ELM WDs}
A UV spectrum has been analyzed   for only one other ELM WD  so far, namely that of GALEX J1717+6757 \citep{her14}. This star has log $g$ = 5.67, essentially identical to that of \hd, and \teff\ = 14,900 K, which gives a mass of $\sim$0.19 \msun.
\citet{her14} detected lines from nine metals, and for all of them they derived an abundance higher (by rougly 0.5 to 2.0 dex) than that measured for \hd. In their discussion, \citet{her14} point out that radiative levitation for N, O, Si, P, and Fe gives predictions that are in line with the abundance pattern seen in the star.  One noteworthy exception is carbon for which strong radiative support was predicted, but the element is extremely underabundant (by a factor of 1000 with respect to the sun) in the atmosphere of J1717. Interestingly, this is reminiscent of the situation for \hd, where important radiative support is also predicted for carbon at 21,000 K, but the element is strongly depleted in the stellar atmosphere. \citet{her14} argue that a possible explanation for carbon depletion might be a history of repeated CNO flashes, as predicted by evolutionary models with $M \ga$ 0.18~\msun. This might also apply to \hd, although we do not see the metal enrichment that might be associated with CNO-flash surface convection.
Calcium is another element that defies radiative levitation predictions in J1717: more Ca is observed than predicted, just as it is for \hd. According to Fig. 5 in \citet{her14} the abundance pattern of J1717 is not very different from those of WDs with ongoing accretion from debris discs, except for Mg and O. However, the authors conclude that debris-disk accretion is unlikely in the case of J1717 because of the discrepancy in oxygen abundance, and because accretion from a circumbinary disk is dynamically difficult to explain.

Two ELM WDs are known to be metal rich: PSR J1816+4510 (\teff\ = 16,000 K, log $g$ = 4.9), the companion of a massive millisecond pulsar \citep{kap13}, and SDSS J0745+1949 (\teff\= 8,380 K, log $g$ = 6.2), a tidally distorted star \citep{gia14b}. The analyses of their optical spectra allowed for the abundance determination of a few elements. The abundances of Mg, Ca, Ti, Cr, and Fe turned out to be close to solar in the case of J0745 while He, Mg, Ca, Si, and Fe are about 10 times solar in  J1816. The origin of the metals in these two stars is not fully understood. A recent shell flash is a possible explanation for the high metallicity of J1816, but the rather low mass inferred for J0745 (M = 0.16 \msun) does not suggest the star could have gone through such shell flash episodes. On the other hand, accretion of metals from a debris disk was discussed as a possibility for J0745, while the pulsar companion of J1816 prevents such an accretion.

Even though detailed abundances are known for only three other ELM-WDs, \citet{gia14} measured abundances of He, Ca, and Mg in their sample of 61 stars (though such elements were visible only in a minority of stars). As mentioned previously, they noticed the presence of Ca lines in basically all of their stars having log $g$ $\la$ 6.0. \hd, which has low gravity, now also falls into this category thanks to the tiny line we identified in its optical spectra. The abundance we derived (log $N$(Ca)/$N$(H) = $-$7.3), though below the \citet{gia14} detection threshold, is comparable to the lowest abundances measured in their cooler ELM WDs (we note here that the \ion{Ca}{ii} lines get stronger at lower temperature). The same observation holds for Mg, where the abundance of \hd\ is similar to the lowest values detected in the \citet{gia14} sample.

\subsubsection{Comparison with sdB stars}
Given its fundamental parameters, \hd\ is considered an sdB star, so it might be worth comparing its abundances with those of sdBs. To this end we also included in Fig. \ref{comp} the ``upper averages'' (including determinations and upper limits) of the \citet{geier13} sample, consisting of a hundred sdBs. 
It thus appears that \hd\ is a particularly metal-poor sdB, but  also an especially cool one. 
%However, one must keep in mind that \citet{geier13} analysis is based on optical spectra and a star depleted in metals like \hd\ would have led to upper limits determination for most of the elements. 
PG1627+017 is a cool pulsating sdB (\teff\ $\sim$21$-$24 kK) that can be compared to \hd\ in terms of effective temperature, but the lower log $g$ of PG1627+017 places it onto the EHB, meaning that the star is likely to be more massive than \hd. Indeed, a mass of $\sim$0.45 \msun\ was suggested by its seismic properties \citep{ran06}. The UV spectrum of PG1627+017 has allowed  a detailed abundance determination, and the star is clearly more metal-rich than \hd\ (see Fig. 6 in \citealt{bla08}). 
Carbon is the most extremely depleted element in \hd. Past studies of sdBs have shown a large star-to-star variation in atmospheric carbon. Very low carbon upper limits (log $N$(C)/$N$(C$_{\odot}$)) = $-$5.7 and $-$5.3) were determined for CD $-$24$\degr$731 and PG 1219+534, respectively \citep{otoole06}. The sdOB star Feige 110 is another example of a carbon poor star \citep{rauch14,heb84}.
Lead is the most enriched element in HD188112, and it has been identified in 22 out of the 27 hot subdwarf UV spectra examined by \citet{otoole04}. Further abundance analysis of sdBs showed that lead enrichment varies from 10 to 1000 times solar \citep{otoole06,bla08}. These two studies also showed that trans-iron elements tend to be overabundant in the atmosphere of sdB stars. In this respect, the abundance pattern of \hd\ is similar to other sdBs.

%\citet{otoole04} looked for heavy elements in the UV spectra of 27 hot subdwarf stars and could identify lead lines in 22 of them as for the 5 remaining stars, the presence of lead remained uncertain. Nevertheless this study pointed out that this element is commonly seen in such stars. Abundances were measured in a dozen sdBs and appears to be enriched in all of them. 

\subsection{Binary system}

In addition to the spectroscopic analysis, we attempted a better characterization of the \hd\ binary system. By combining our new UV radial velocity measurement with the already published values, we refined the parameters of the binary system: period (0.60658584(5) d), amplitude (188.7 $\pm$ 0.2 km s$^{-1}$), and systemic velocity (26.6 $\pm$ 0.2 km s$^{-1}$).
 Access to a Hipparcos parallax for a star provides an independent mass measurement, i.e., independent of stellar evolution models.
In our case, the spectroscopic mass of 0.245~$^{+0.075}_{-0.055}$ \msun\ derived with the parallax distance agrees quite well with that indicated by evolutionary models of low-mass He-core WDs, which predict 0.211 $\pm$ 0.0175 \msun\ \citep{alt13}. However, the uncertainty on the spectroscopic mass is rather large (due to the uncertainty on the parallax itself) and allows for a mass up to 0.32 \msun. Such a mass is at the lower limit of He-core burning for an sdB star  \citep{han02,hu07}. Low-mass sdB stars are created when a  massive progenitor (2$-$2.5 \msun) ignites helium under non-degenerate conditions. According to binary synthesis models, the second common envelope ejection channel can produce sdBs with masses down to 0.33~\msun\ in close binaries with a WD companion \citep{han03}. The position of \hd\ in the log $g$ - \teff\ diagram would fit within the 0.33 \msun\ EHB band (see Fig. 7 from \citealt{sil12}). 
Although there is a possibility that \hd\ is a He-core burning object, one must keep in mind that these low-mass sdBs occur much less frequently than the canonical ($\sim$0.47 \msun) sdBs and that the extreme upper limit of our spectroscopic mass coincides with the extreme lower limit of the He-core burning. Given the current parallax value, we thus consider it unlikely that \hd\ is such an object. 
For the lower mass limit on \hd, both the parallax and evolutionary models give a value of $\sim$0.19 \msun. In this case, the star is certainly a pre-ELM WD.

In our attempt to constrain the mass of the WD companion, we considered both the 0.21 and 0.245 \msun\ solutions indicated respectively by evolutionary models and the parallax. We derived a minimum mass of 0.7 \msun\ for the companion\footnote{If we consider  0.19~\msun\ for \hd\, the minimum mass of the companion decreases only slightly to 0.68~\msun.}. 
It is interesting to compare this minimum mass with the minimum mass distributions of sdB binaries with WD companions and of ELM WD binaries illustrated in Fig. 14 in \citet{kup15}. When considering the distribution for sdB binaries, it appears that a 0.7 \msun\ companion is quite massive; only two known sdBs have a more massive WD companion\footnote{\hd\ was not included in the \citet{kup15} sample because of its low mass.}. After a survey was dedicated to finding massive companions to sdBs \citep{geier11}, it now seems that such systems are quite rare. 
On the other hand, when looking at the observed minimum mass distribution of ELM systems (data from \citealt{gia14}), the companion masses are  evenly distributed between 0.1 and 0.9 \msun\ with a few more massive outliers. In this context, the companion of  \hd \   would be on the massive side, but not outstandingly so. 

We measured the rotational broadening of metallic lines in our UV spectra and derived $v_{\rm rot}$ sin $i$ = 7.9 $\pm$ 0.3 km s$^{-1}$. In order to constrain the inclination of the system, we assumed it to be tidally locked, which led to inclination values between 49$\degr$ and 61$\degr$. This range nicely overlaps with the most likely inclinations for a random sample. 
Our inclination range indicates companion masses between 0.92 and 1.33~\msun. The synchronization assumption thus seems reasonable, or at least does not lead to improbable results.
Recent statistical investigations on the mass distribution of companions to ELM WDs found a mean mass of $\sim$0.75 \msun, with $\sigma$ $\sim$0.25 \msun\ \citep{and14,bof15}. Both studies found very similar results when assuming a Gaussian distribution to model the companion's mass population and comparing it with the observed ELM WD systems of \citet{browr13} and \citet{gia14}. If the \hd\ system is indeed synchronized, the companion would thus be on the massive side of these mass distributions. 
The possibility of a neutron star companion should not be discarded, as it would require an inclination angle only slightly below 49$\degr$, and would imply the star is rotating faster than synchronized. This last assumption is plausible considering the fact that evolutionary models predict that the rotation of the star should increase after the ejection of the common envelope, as the star is contracting toward the white dwarf regime. 
Given the orbital parameters of the \hd\ system, the rotation period of the star can hardly be less than half the orbital period, in which case the inclination would be less than 25$\degr$ and the companion mass above 6 \msun. A search for possible X-ray emission from the companion of \hd\ might help to constrain its nature \citep{kil14}.

It might be interesting to look at the possible future of the system. 
As the star contracts toward the cooling track it is likely to experience hydrogen shell flashes 
during which the star will expand and fill its Roche lobe \citep{ist14}. Brief episodes of mass transfer might lead to mass loss of a few times 10$^{-4}$
M$_\odot$ during a shell flash (see Fig. 12 in \citealt{nel04}). 
Because the hydrogen fuel is diminished both by burning and mass loss, the evolution of the star will speed up.
In the long run, gravitational wave radiation will lead to a decay of the orbit resulting finally in a merger of the components after a few Hubble times.  
By the time of merging, the system has evolved into a degenerate He-core white dwarf and a massive companion, most likely a white dwarf of 0.7 to 1.3 M$_\odot$. 

The outcome of the merger depends strongly on the mass of the white dwarf companion and its composition. If the mass is as low as 0.7  M$_\odot$, the white dwarf is of C/O composition and the merger most likely produces a RCrB star \citep{web84,clay07}. 
Mergers of He-core white dwarfs with more massive C/O white dwarfs may lead to SNe Ia \citep{pak13,dan15}. Some simulations showed that the He-shell ignition induced by the accretion of material onto a sub-Chandrasekhar C/O WD is very likely to trigger a detonation in the core, leading to a SN Ia explosion (double-detonation scenario, \citealt{fink07,fink10,moll13}). If the core fails to ignite\footnote{In the case of a very low-mass (and low-density) C/O WD companion or a high-mass WD companion with O/Ne composition, a secondary detonation may be less likely \citep{shen14}.},
 then the helium explosion alone can produce a fainter type .Ia SN \citep{bild07}.
%However, models for component masses appropriate are lacking.
At the high end of the companion mass range ($>$1.1 M$_\odot$), the system would have a low mass ratio ($q$ $\la$0.20) and thus evolve via stable mass transfer into an AM Canum Venaticorum (AM CVn) \citep{marsh04, dan11}, which is a potential SN Ia progenitor \citep{shen14}.
However, it is usually assumed that white dwarfs more massive than about 1.1 M$_\odot$ have O/Ne/Mg cores and these objects, when accreting matter, are not likely to explode, but would rather collapse into a neutron star via electron capture \citep{nom91,shen14}.
On the other hand, recent investigations showed that near-Chandrasekhar mass ($\sim$1.1$-$1.3 \msun) white dwarfs may have a hybrid composition of C/O/Ne and could be prone to explode, leading to faint SNe of the observed type Iax \citep{chen14,kro15}.
In either case, the outcome of the system depends on the stability of the mass transfer, which in turn depends on the mass ratio of the components \citep{marsh04}. For mass ratio $q$ in the range $\sim$0.25--0.70 the mass transfer can be either stable or unstable (see Fig. 1 in \citealt{dan11}). For the \hd\ system, $q$ can be up to $\sim$0.32 if the companion mass is close to the minimum value.

\section{Conclusion}

We began our analysis of the UV spectra of \hd\ with the main goal of determining the projected rotational velocity, in order to put tighter constraints on the mass of the companion. However, the high-resolution spectra showed a large number of metallic lines that allowed us to perform a  comprehensive abundance analysis of this pre-ELM WD. This is extremely interesting since the chemical compositions of only a few ELM WDs have been determined to date. Additionally, because it is  a pre-ELM WD, \hd\ represents an earlier evolutionary stage and comparing its metallic content with those of cooler, more compact, and thus older (in terms of cooling age) ELM WDs might help to determine the origin of their metals. 

%\textbf{NEW}
The spectral analysis was performed using hybrid NLTE model atmospheres, in which the NLTE effects are taken into account in the computation of the population numbers while the atmospheric structure is computed in LTE. The computation of NLTE populations was only possible for a limited number of species (C, N, O, Mg, Al, Si, S, Ca, and Fe); otherwise, the LTE approximation had to be used. By comparing both LTE and NLTE spectra for a few metallic species, we concluded that the spectral lines originating from the main ionization stages of an element were not affected, or were only marginally affected, by NLTE effects. Thus, they can be used reliably to derive abundances with LTE models. However, we clearly saw in our model spectra that NLTE effects affect lines of non-dominant ionic species. Fits of these lines led to discrepant abundances, by up to $\sim$0.5 dex. As these effects are seen in \hd, they are likely to also be present in stars having a temperature higher than $\sim$21,000 K. 
This is why we emphasize that it is more accurate to consider only lines originating from the dominant ionic species when using LTE models in order to derive appropriate abundances. 

Based on our abundance determination and comparisons with other ELM WDs and hot subdwarf stars, \hd\ appears to be an especially metal-poor object. Its abundance pattern suggests diffusion effects where gravitational settling is probably counteracted by radiative levitation. In constrast to other more metal-rich ELM WDs, \hd\ does not require any ``metal enrichment'' mechanism to explain the presence of metals in its atmosphere. It is also unlikely that its metallicity will increase as it ultimately cools down and becomes more compact; radiative levitation should become weaker as \teff\ decreases and gravitational settling stronger as log $g$ increases. However, given its mass, \hd\ is likely to experience an  H-shell flash that would temporarily increase the surface abundance of metals, especially He and N, although gravitational settling is expected to act quickly as the star contracts again.
Considering the four detailed abundance analyses currently available  for ELM/pre-ELM WDs, it seems that they show a rather wide variety of metallicities, from metal-rich to metal-poor, as is also observed for sdB stars.

Despite our efforts to constrain the mass of the \hd\ companion, we are limited by the unknown inclination of the system and by the uncertainty on the mass of \hd\ itself.
By assuming tidally locked rotation, we could constrain the inclination to be between 49$\degr$ and 61$\degr$. Unfortunately, we cannot currently test or verify whether the system is indeed tidally locked. Nevertheless, this assumption leads to inclinations close to the most likely ones for a random sample, so it does not appear to be inappropriate. This inclination range indicates a companion mass between 0.92 and 1.33~\msun, i.e., a rather massive WD, while the minimum mass is 0.7 \msun.
For the mass of \hd,\  a value up to 0.32 \msun\ is possible according to the Hipparcos distance, in which case the star would be at the limit of the He-burning mass range. Gaia should provide us with a more precise distance.

\begin{acknowledgements}
This work was supported by the the Deutsches Zentrum f\"{u}r Luft- und Raumfahrt (grant 50 OR 1315). Based on observations made with the NASA/ESA Hubble Space Telescope, obtained at the Space Telescope Science Institute, which is operated by the Association of Universities for Research in Astronomy, Inc., under NASA contract NAS 5-26555. These observations are associated with program no. 12865, cycle 20. We are most grateful to P. Chayer for sharing the results of his radiative levitation computation and T. Rauch for fitting the tin lines with TMAP model atmospheres. We also thank A. Istrate and H. Dreschel for interesting discussions. The work of FKR is supported by the ARCHES prize of the German Ministry of Education and Research (BMBF) and by the DAAD/Go8 German-Australian exchange programme for travel support. WH and ST acknowledge support by project TRR 33
'The Dark Universe' of the German Research Foundation (DFG)
% and to H. Dreschel for providing us with an upper limit for the eccentricity. 
This research has made use of ISIS functions provided by ECAP/Remeis observatory and MIT (http://www.sternwarte.uni-erlangen.de/isis/). We thank John E.\ Davis for the development of the {\sc slxfig} module used to prepare some of the figures in this paper.
\end{acknowledgements}

%\clearpage
% References
%
\bibliographystyle{aa}
%\bibliography{reference}

\end{document}